\documentclass[twocolumn,aps]{revtex4-1}
\usepackage{graphicx}
\usepackage{dcolumn}
\usepackage{bm}
\usepackage{color}
\usepackage[english]{babel}

\newcommand{\dnua}{0.1}
\newcommand{\dnub}{0.05}
\newcommand{\dtau}{20}

\begin{document}

\title{Direct observation of Rogue Waves in optical turbulence  using  Time Microscopy}  

\author{Pierre Suret}
\email[Corresponding author : ]{Pierre.Suret@univ-lille1.fr}

\author{Rebecca El Koussaifi}
\author{Alexey Tikan}
\author{Cl\'ement Evain}
\author{St\'ephane Randoux}
\author{Christophe  Szwaj} 
\author{Serge Bielawski}

\affiliation{Laboratoire de Physique des Lasers, Atomes et Molecules,
  UMR-CNRS 8523,  Universit\'e de Lille, France \\
Centre d'Etudes et de Recherches Lasers et Applications (CERLA), 59655 Villeneuve d'Ascq, France}

\maketitle

{\bf The formation of coherent structures in noise driven
  phenomena and in Turbulence is a complex and fundamental
  question~\cite{Sirovich:87}. A particulary important structure is
  the so-called Rogue Wave (RW) that arises as the sudden appearance
  of a localized and giant peak ~\cite{Onorato:01,Pelinovskybook2,
    Onorato:13}. First studied in Oceanography, RWs have been
  extensively investigated in Optics since 2007~\cite{Solli:07}, in
  particular in optical fibers experiments on supercontinua
  ~\cite{Solli:07, Erkintalo:09, Mussot:09} and optical turbulence
  ~\cite{Walczak:15,Onorato:13}.  However the typical timescales
  underlying the random dynamics in those experiments prevented --up
  to now-- the direct observation of isolated RWs. Here we report on
  the direct observation of RWs, using an ultrafast acquisition system
  equivalent to microscope in the time domain \cite{Kolner:89,Bennett:99,  Foster:08,Okawachi:12}. The RWs are generated by nonlinear propagation of
  random waves inside an optical fiber, and recorded with $\sim
  250$~fs resolution. Our experiments demonstrate the central role
  played by ``breathers-like'' solutions of the one-dimensional
  nonlinear Schr\"odinger equation (1D-NLSE) in the formation of RWs
  ~\cite{Akhmediev:09}.}

Common oceans waves are weakly nonlinear random objects having nearly
Gaussian statistics, while at the same time, RWs are  waves of
extremely large amplitude that occur more frequently than expected from the normal law
\cite{Onorato:01, Pelinovskybook2, Onorato:13}. The mechanisms
underlying the generation of coherent structures 
such as RWs from the nonlinear propagation of  random waves i.e., in turbulent flows, is
a subject of very active debates and still represents an open question
\cite{Hammani:10, Walczak:15,Agafontsev:15,Toenger:15}.

From the theoretical point of view, the so-called {\it focusing} 1D-NLSE (see
Eq. \ref{eq:NLS1D}), which is a generic equation  having an ubiquitous importance in Physics, plays 
a central role in this debate
\cite{Onorato:01, Onorato:13, Akhmediev:13, Dudley:14}. In
particular, the 1D-NLSE describes at leading order the physics of
deep-water wave trains and nonlinear propagation in optical fibers \cite{Chabchoub:15}. The breather-like solutions of the 1D-NLSE, also called solitons on finite background, are  now considered as being prototypes of RWs \cite{Akhmediev:09, Akhmediev:13,Kibler:10,Chabchoub:11,Kibler:12}.

Despite the numerous experimental works devoted to optical RWs,
\cite{Solli:07, Erkintalo:09, Mussot:09, Hammani:08, Walczak:15,
  Onorato:13, Dudley:14} the direct observation of these coherent
structures in the time domain has never been reported in the context
of the nonlinear propagation of  {\it random waves}. Randomness of the
initial condition is known to play a crucial role in the generation of
RWs as it has been pointed out in the supercontinuum driven by noise
~\cite{Solli:07,  Mussot:09, Erkintalo:09, Dudley:14} or in optical
turbulence \cite{Walczak:15,  Onorato:13, Dudley:14, Picozzi:14}

Contrary to the experiments performed in the spatial domain
\cite{Bromberg:10, Pierangeli:15}, the fast time scales of
fluctuations (picoseconds or less) involved in single-mode fiber
experiments makes single-shot recording of RWs a particularly
challenging task. Pioneer works hence naturally provided {\it indirect
  evidences} of RWs, using e.g. spectral filtering \cite{Solli:07,
  Erkintalo:09, Mussot:09} or statistical measurement from optical sampling techniques~\cite{Walczak:15}.\\

\begin{figure*}
\includegraphics[width=17cm]{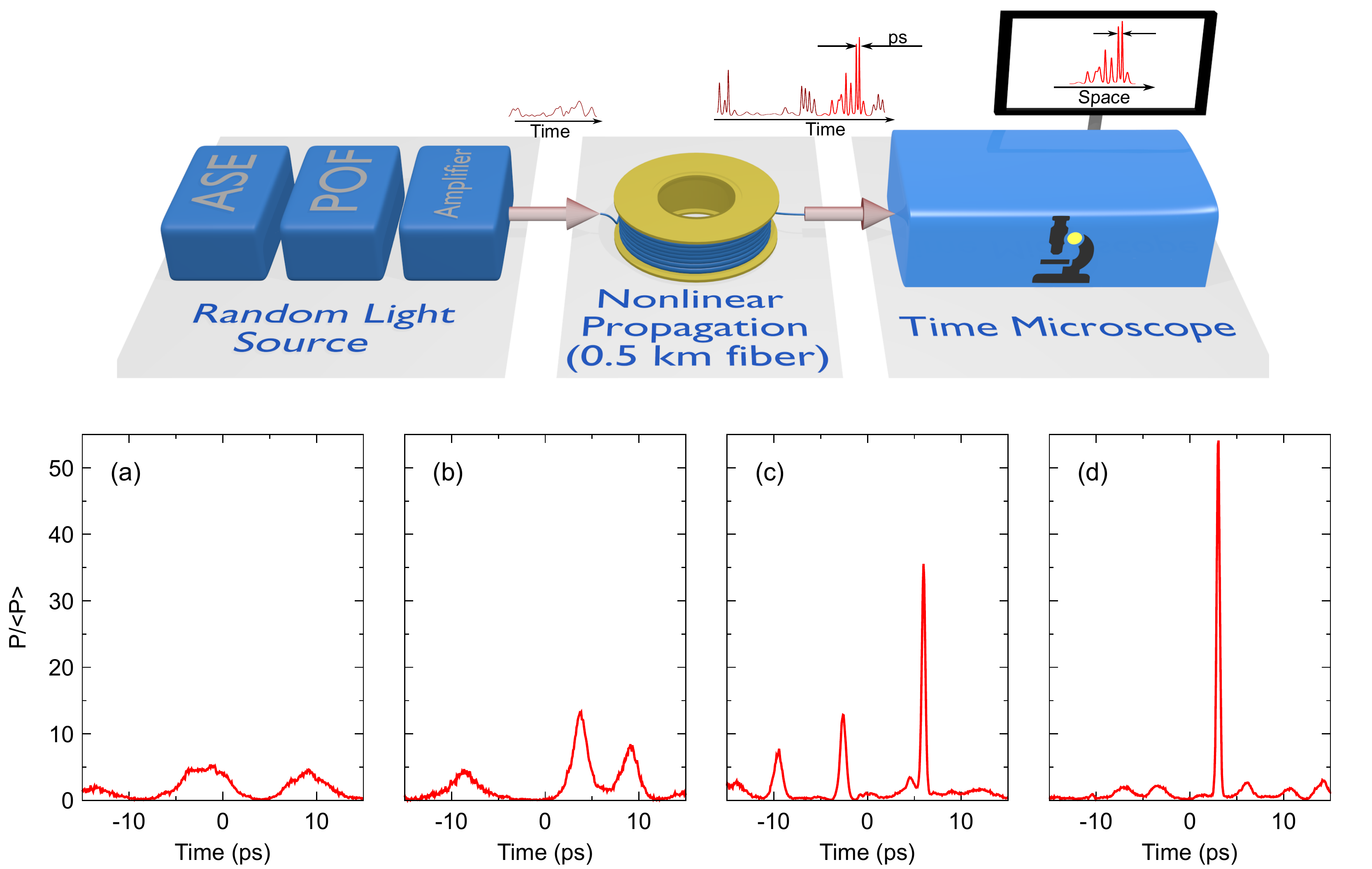}
\caption{{\bf Overview of Experiments} (Upper part) Global strategy
  for the experimental observation of optical rogue waves. Incoherent
  light from an 1560~nm Amplified Spontaneous Emission source (ASE) is
  filtered by using a Programmable Optical Filter (POF) and amplified before
  experiencing nonlinear propagation in a polarization maintaining
  (PM) fiber.  Random fluctuations of the output power are analyzed in
  single-shot with sub-picosecond resolution, using a specially
  designed time-microscope, which maps the temporal evolution onto the
  spatial coordinate of a conventional camera (see Fig~\ref{fig:setup}
  and Methods for details). (a-d) Typical single-shot recordings of
  random waves. Initial spectral width $\Delta \nu=0.1$THz. (a)
  Initial condition. (b-d) RWs observed at the output of the fiber
  for mean powers $\langle P \rangle=0.5$W (b), $\langle P
  \rangle=2.6$W (c), $\langle P \rangle=4.$W (d)}
\label{fig:1}
\end{figure*}

In this Letter, we present direct single-shot recordings of optical
RWs by using a specially-designed {\it Time Microscope} (TM) ultrafast
acquisition system \cite{Kolner:89,Bennett:99, Okawachi:12}. The
temporal resolution of $\sim$250~fs of our TM (see Methods and
supplementary material) allows us to investigate the fast dynamics
arising from the nonlinear propagation of random waves in an optical
fiber (upper part of Fig.~\ref{fig:1}). Observations performed with the
TM at the output of the fiber immediately reveals the emergence of
intense peaks, with powers frequently exceeding the average power
$\langle P \rangle$ by factors of 10-50 [see Fig~\ref{fig:1}(b-d) and
supplementary movie 1].  Starting from random fluctuations having
typical time scale around $5-10$ps [Fig~\ref{fig:1}(a)], those extreme
peaks are also found to be extremely narrow, with
time scales of the order of several hundreds of femtoseconds
[Fig~\ref{fig:1}(b-d) and Fig~\ref{fig:dynamics}(b-d)].\\

\begin{figure*}
\includegraphics[width=17cm]{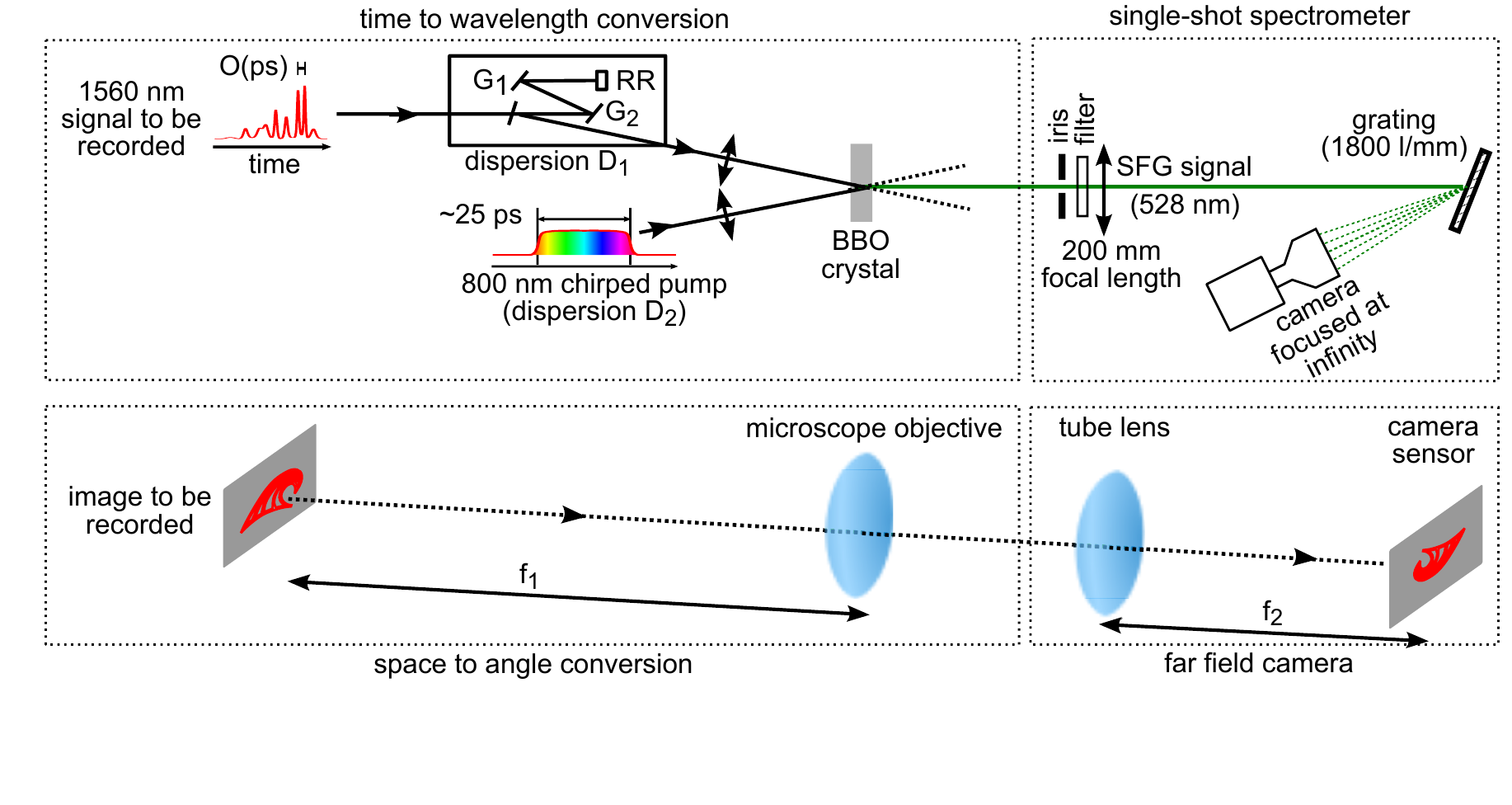}
\caption{{\bf Time microscope realized for the ultrafast acquisition
    of rogue waves}. Upper part: experimental setup. A key element is
  the time lens, which is composed by the BBO crystal, pumped by the
  stretched 800~nm pulse. Lower part: spatial analog of the time
  microscope. The dispersion $D_1$ (provided by the grating
  compressor) is analog to the diffraction between the initial image
  and the lens with focal length $f_1$. The time lens is the analog of
  the ($f_1$) lens. The single-shot spectrometer is formally analog to
  the far-field camera. G$_1$ and G$_2$ are 600~l/mm gratings, RR is a
  roof retroreflector. Note that the BBO crystal is placed at the
  focal plane of the 200~mm collimating lens. Transport optics are not
  represented.}
\label{fig:setup}
\end{figure*}

\begin{figure*}
\includegraphics[width=17cm]{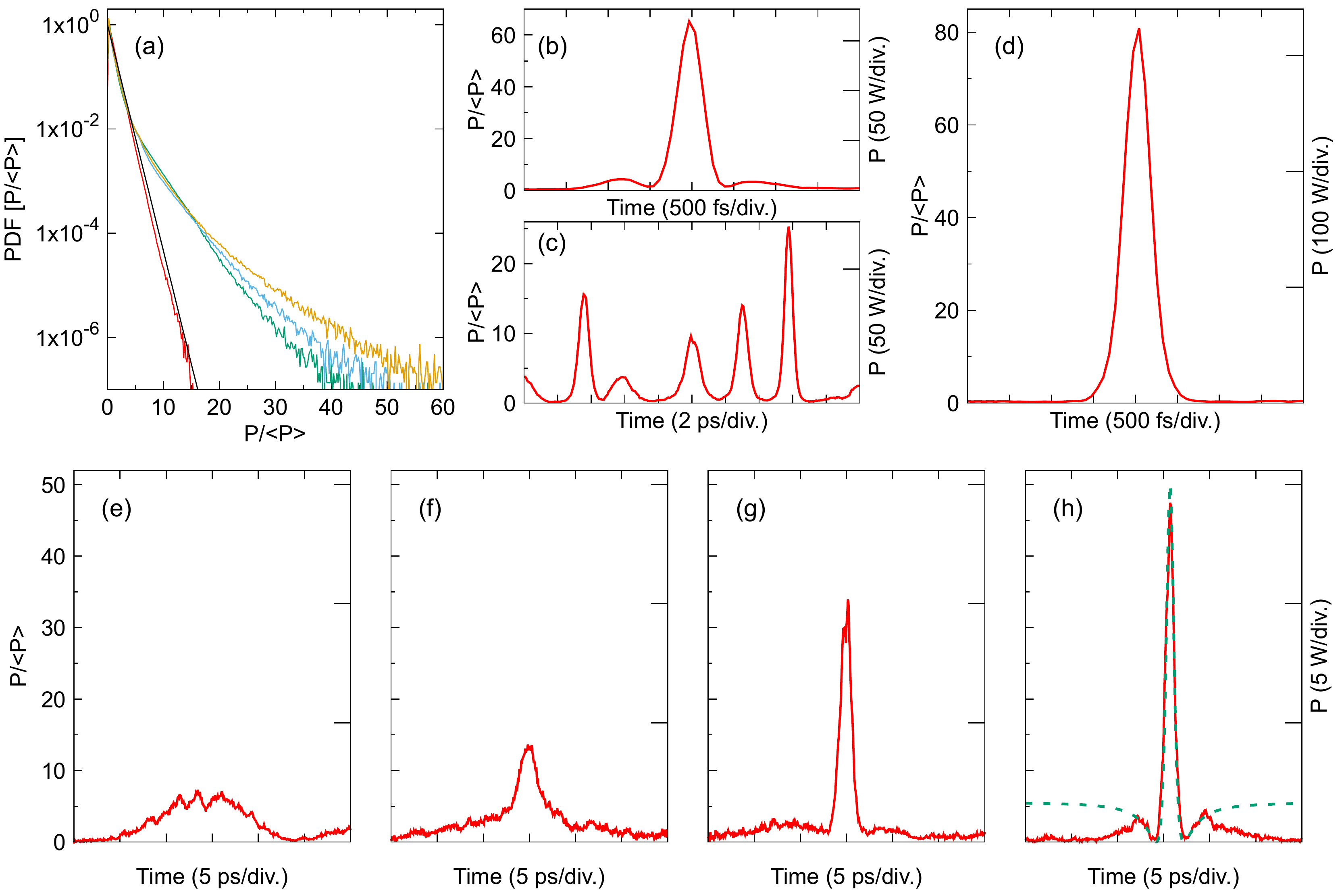}
\caption{{\bf Typical power profiles and Statistics of Rogue Waves (Experiments)}
(a) PDF of the initial condition (red line) and of the output power at $500$ mW 
(green line), $2.6$ W (blue line), $4$ W (yellow line). The black line represents 
the normalized exponential distribution : $\mathrm{PDF}(P/\langle P\rangle)=\exp(-P/\langle P\rangle)$. (b-d) Typical
random signals recorded by the TM at the output of the nonlinear PM fiber. (b) PS-like 
structure ($P=2.6$ W, $\Delta \nu=0.1$ THz). (c) Series of multiple peaks 
($P=2.6$ W, $\Delta \nu=0.1$ THz). (d) Giant optical RW reaching a peak power $\sim 80$
times greater than the mean power of the random wave ($P=4$ W, $\Delta \nu=0.1$ THz). (e-h) Typical snapshots 
showing several stages leading to the emergence of a breather-like structure ($P=300$ mW, $\Delta \nu=0.05$ THz). 
The green dashed line in (h) represents the best fit to the analytical expression of a PS.}
\label{fig:dynamics}
\end{figure*}

More precisely, the random waves used as initial conditions in our
experiments are partially coherent light waves emitted by a high power
Amplified Spontaneous Emission (ASE) light source at a wavelength
$\lambda\sim1560$ nm (see Fig.~\ref{fig:1}).  Using a programmable
optical filter, the optical spectrum of the partially coherent light is precisely designed
to assume a Gaussian shape having a full width at half maximum that is
adjusted either to $\Delta \nu =$\dnua THz or $\Delta \nu =$\dnub
THz. The partially coherent waves are launched into a 500~m-long
single mode polarization maintaining fiber at a wavelength falling
into the anomalous (focusing) regime of dispersion.  The light at the
output of the nonlinear fiber is then directed to the TM (detailed in
Figure~\ref{fig:setup}), which acquires traces (optical power versus
time over a $\approx \dtau$~ps-long window) at a rate of 500 per
second, and displays the signals in real time.

As in a standard spatial imaging microscope, the TM is composed of an
objective and a tube lens [see Fig.~\ref{fig:setup}.(b)]. The
objective is a time lens \cite{Kolner:89,Bennett:99, Okawachi:12}
operating from sum-frequency generation (SFG) between the 1560~nm
signal and a chirped pump pulse (at 800~nm) (see Sec. Methods). The
observation in the focal plane of the tube lens is achieved by a
spectrum analyser (composed of a diffraction grating, a lens and a
camera). {The use of this TM strategy enables to easily reach extremely high dynamical ranges (up to $40$~dB, see
Methods), which is a crucial point for analyzing extreme events embedded in
moderate power fluctuations.}

In order to quantify the emergence of RWs, we compute statistical
distributions from a large amount of data recorded with the TM at the
input and at the output ends of the fiber (see Sec. Methods). As expected,
the probability density function (PDF) of the optical power emitted by
the ASE source is systematically very close to the exponential
distribution that correspond to a Gaussian statistics for the field
[see Fig.~\ref{fig:dynamics}.(a)]~\cite{Goodman:85,
  Walczak:15,Agafontsev:15}.  On the contrary,  the PDF of light power
at the output of the nonlinear fiber is found to exhibit heavy-tailed
deviations from the exponential distribution, thus confirming the
generation of RWs [see Fig. \ref{fig:dynamics}.(a)].  Moreover the
PDFs demonstrate that the number of RWs having high peak power increases while the mean power $\langle P \rangle$ of random optical waves  ({\it i.e.} the strength of nonlinearity) increases  [see Fig. \ref{fig:dynamics}.(a)].

To the best of our knowledge, experimental signals plotted in
Figs.~\ref{fig:1}.(a-d) and ~\ref{fig:dynamics}.(b-h) represent the
first direct and accurate observation of the RWs underlying these
heavy tailed statistics. Starting from random light propagating with a
mean power of 4 W in the fiber, huge RWs having peak power that exceeds
300 W can be observed at the output of the fiber [Fig.\ref{fig:dynamics}.(d)~]. From a careful analysis of the data,  two
typical shapes can be distinguished  : isolated breather-like RWs [see
Fig. \ref{fig:dynamics}.(b)] and more complicated structures composed of several peaks
[see Fig. \ref{fig:dynamics}.(c)]. In order to illustrate the process of emergence of
breather like RWs, we plot in  Fig. \ref{fig:dynamics}.(e-h) different
structures observed after the propagation of partially coherent waves
having an initial spectral width $\Delta \nu = \dnub$THz and an
average power  $\langle P \rangle=  0.3$W.  We have selected these
structures because their shapes are strikingly similar to those found in the
scenario leading to the formation of the  Peregrine soliton (PS) while
starting from a single hump at initial stage \cite{Bertola:13}.
Remarkably, the {\it power} profile of the exact analytical PS
coincide very well with experimental structures [see green dashed line
in  \ref{fig:dynamics}.(h)]. However note that the precise
identification of the breathers-like structures (as PS or Akhmediev
breather or other more complex solutions of 1D-NLSE) would require a
simultaneous measurement of the phase dynamics~\cite{Randoux:15}.\\

Behaviors observed in experiments can be well reproduced from numerical simulations
of the 1D-NLSE (see Sec. Methods). First of all, the PDFs of optical power
[see Fig.~\ref{fig:num}.(a)] and the optical spectra (see Supplementary Material)
are well reproduced by numerical simulations. Fig. \ref{fig:num} shows a picture
of typical random fluctuations of the optical power that are found at the input and
output ends of the optical fiber. Taking a partially-coherent
light field having a bandwidth of $\dnua$ THz at initial stage, the
typical time scale for power fluctuations is around a few picoseconds
[Fig. \ref{fig:num}(b)]. The scenari observed in the experiments  are
also found in numerical simulations. In particular,  either breather-like structures appear and
disappear along the propagation [see Fig.\ref{fig:num}(g)], either
several pulses simultaneously emerge together from the random
background [see Fig. \ref{fig:num}(c)]. Our experiments provide
snapshots randomly recorded while the numerical simulations allow to
follow the dynamics of nonlinear random waves along the propagation. In this respect, the numerical simulations reveal that the breather-like structures
often emerge on the top of the initial power fluctuations (see Fig. \ref{fig:num}(d-g)
and video in Supplementary Material).\\

 \begin{figure*}
\includegraphics[width=17cm]{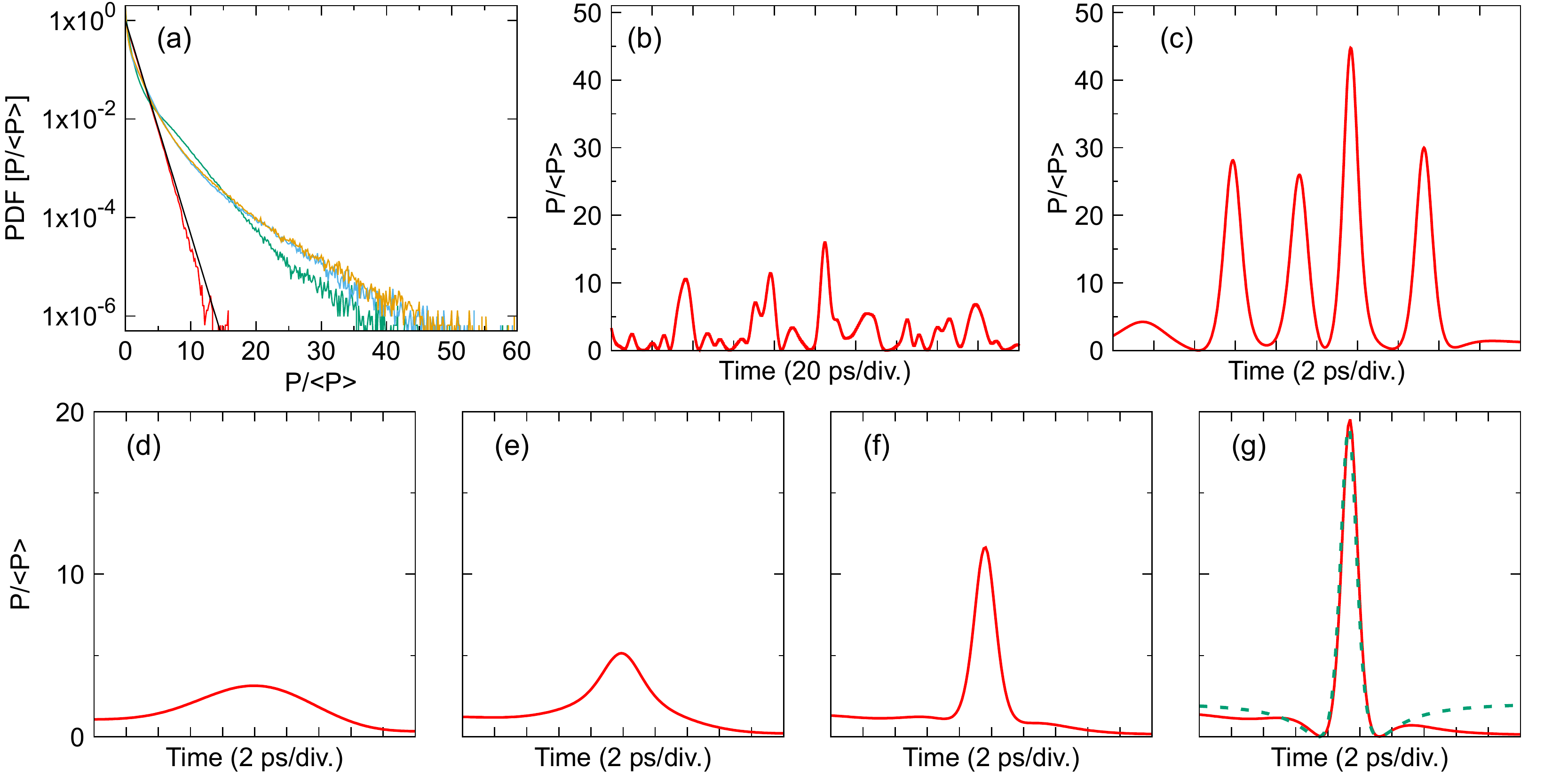}
\caption{{\bf Typical power profiles and Statistics of Rogue Waves (Numerical Simulations of the 1D-NLSE)}
(a) PDF of the initial condition (red line) and of the output power at $500$ mW
(green line), $2.6$ W (yellow line), $4$ W (blue line). The black line represents
the normalized exponential distribution :  $\mathrm{PDF}(P/\langle P\rangle)=\exp(-P/\langle P\rangle)$. (b) Typical
fluctuations of the power of the random field used as initial condition. (c) Series of multiple peaks (zoom) ($P=2.6$ W, $\Delta \nu=0.1$ THz).
(d-h) Typical snapshots (zoom) showing several stages leading to the emergence of a breather-like structure along the propagation ($P=1$ W, $\Delta \nu=0.05$ THz, $z=0$ (d), $z=285$ (e), $z=415$m, (f) $z=500$ m ). The green dashed line in (g) represents the best fit to the analytical expression of a PS.}
\label{fig:num}
\end{figure*}

 In the last years, the common and shared conjecture is that
 breather-like solutions of 1D-NLSE such as PS or Akhmediev breathers
 represent  prototypes of RWs  \cite{Akhmediev:09,
   Akhmediev:13,Kibler:10,Chabchoub:11,Kibler:12,
   Dudley:14,Toenger:15}. This has motivated very nice experiments  in
 which these solitons on finite background  have been generated in a
 deterministic way in optical fibers \cite{Kibler:10,Kibler:12,
   Frisquet:13} and in  a one-dimensional water tank
 \cite{Chabchoub:11}. These experiments  make use of
 carefully-designed  {\it coherent initial conditions}.

 On the contrary, the initial conditions in our experiments are
 designed to be ``ocean-like'' {\it random
   waves}~\cite{Onorato:13, Janssen:03}. In this context of nonlinear propagation
 of random waves, previous experimental works performed in a 1D water
 tank~\cite{Onorato:04} and in optical fibers~\cite{Walczak:15} have
 revealed heavy-tailed deviations from Gaussian statistics. For the
 first time, our time-resolved observations correlate in an
 unambiguous way the occurrence of this heavy-tailed statistics with
 the frequent occurence of breather-like coherent structures. Our
 experimental observations favor a well-known scenario in which a PS
 emerges from a single real hump \cite{Bertola:13}. However it must be
 emphasized that the precise identification of coherent structures
 requires the precise knowledge of the phase evolution.  The
 simultaneaous fast measurement of phase and amplitude fluctuations
 therefore represents the next experimental bottleneck for the careful identification of optical RWs.\\

 The emergence of coherent structures is a general and mysterious
 feature of stochastically driven processes such as turbulence,
 supercontinuum generation or pattern formation~\cite{Solli:12,
   Dudley:14,Sirovich:87}. The time-resolved direct observation of RWs
 presented in the letter opens the way to numerous studies on the
 relationship betwen coherent structures and noise driven phenomena. In particular, shot by shot {\it spectral}
   measurements in pulsed experiments recently revealed the fascinating
   complexity of the statistical features associated to the so called
   modulation instability ~\cite{Solli:12}. By using our TM, the
   underlying dynamics of the so-called process of noise-driven
   modulational instability and its nonlinear stage is an open
   fundamental question that can now be studied
   ~\cite{Solli:12,Agafontsev:15}.

\section*{Methods}

The partially coherent light (i.e., the initial condition) is
generated by an Erbium fiber broadband Amplified Spontaneous Emission (ASE) source (Highwave),
which is spectrally filtered (with programmable shape and linewidth)
using a programmable optical filter ( Waveshaper 1000S, Finisar). The output is then
amplified by an Erbium-doped fiber amplifier (Keopsys). This
random light is launched into a single-mode polarization maintaining
fiber (Fibercore HB-1550T), with 500~m length and a dispersion
$\beta_2=-20$~ps$^2$km$^{-1}$ (measured). For a given spectral width,
the power of the light launched inside the fiber is controlled using a
half wavelength plate and a polarizing cube.

For the single-shot acquisition of the sub-picosecond optical signals, we
realized an {\it upconversion time-microscope}, largely based on the work of
Ref.~\cite{Bennett:99}. From the input-output point of view, the
time microscope encodes the temporal shape of the optical signal onto the
spectrum of a chirped pulse (i.e., spectral encoding). Then the spectrum is
recorded using a simple spectrometer composed of a 1800~l/mm grating and a
sCMOS camera. A region of interest (of typically 2048x8 pixels) is selected for
recording the image (a raw image is presented in Fig.~2 of the supplementary
material).

For reaching high temporal resolution, a key element is the
time-lens~\cite{Kolner:89}, which is composed by a BBO
crystal, pumped by a chiped 800~nm pulse. Before entering the time
lens, the 1560~nm signal experiences anomalous dispersion in a classic
Treacy grating compressor (see Fig.~\ref{fig:setup}).

As in other time-lens systems~\cite{Kolner:89,Bennett:99, Okawachi:12,  Foster:08}
high resolution requires proper adjustement of the 1560~nm compressor
(see supplemental material for adjustement detail, and performances of
the setup). Conceptually, this is exactly analog to the tuning of the
object-microscope objective distance in classical microscopes. For all
results presented in this paper, the temporal resolution is 250~fs
FWHM, and the field of view is of the order of 20~ps.

As another crucial point, the time-microscope strategy
  leads to an extremely high dynamical range (i.e., the ratio between
  maximal recordable signal and dark noise). This directly stems from
  the choice of employing a camera  for the recording. More precisely, our
  16~bit sCCD camera has an RMS dark noise of $\approx 2$ electrons
  and a saturation value of 30000 electrons, leading to a $\approx
  40$~dB dynamical range.

The 800~nm pump is provided by an amplified Titanium-Sapphire laser (Spectra
Physics Spitfire, 2~mJ, 40~fs, a spectral bandwidth of about 25~nm), operated at 500~Hz,
and only 20~nJ are typically used here. For inducing (normal) dispersion on
the 800~nm pulses we simply adjusted the amplifier's output compressor. The 
dispersion was fixed to $0.23~$ps$^2$, leading to chirped pulses of duration of about $20~$ps.
The 1560~nm grating compressor uses two 600~l/mm gratings, operated at
an angle of incidence of 40~degrees, and whose planes are separated by
42~mm.  The BBO crystal has 8~mm length and is cut for noncollinear
type-I SFG. Focusing of the 800~nm and 1560~nm signals on the BBO
crystal are performed by two lenses with 20~cm focal lengths. In order
to improve the rejection of the 800~nm and the 1560~nm and to keep
only the SFG at $528$~nm, a 40-nm bandpass filter around 531~ nm (FF01
531/40-25 Semrock) is added after the crystal. The camera is a
sCMOS Hamamatsu Orca flash 4.0 V2 (C11440-22U), equiped with a 80~mm lens
(Nikkor Micro 60~mm f/2.8 AF-D). The objective is focused at infinity and the
waist of the SFG in the BBO crystal is imaged on the camera sensor. The camera
is synchronized on the 800~nm laser pulses, and the integration time is
adjusted to 1~ms, thus enabling single-shot operation of the time-microscope.
PDFs of optical power are computed with $75.10^6$ samples ($10^2$ points taken in
the center of the temporal field of the TM from $75.10^4$ frames) for a given set of parameters.

Numerical simulations are performed by integrating the 1D-NLSE :
\begin{equation}
  \label{eq:NLS1D}
  i\frac{\partial \psi}{\partial z}=\frac{\beta_2}{2}\frac{\partial^2 \psi}{\partial t^2}-\gamma|\psi|^2\psi,
\end{equation}
where $\psi$ is complex envelope of the electric field, normalized so that $|\psi|^2$ is the
optical power, $z$ is the longitudinal coordinate in the fiber, and $t$ is the
retarded time. $\beta_2=-20$~ps$^2$km$^{-1}$
is the second-order dispersion coefficient of the fiber and
$\gamma=2$~W$^{-1}$km$^{-1}$ is the Kerr coupling coefficient. All numerical
integrations are performed using an adaptive stepsize pseudospectral method,
using a mesh of 2048 points, over a temporal window of $\Delta T= 250$ps.

In numerical simulations presented in this letter, we neglect linear
losses ($\simeq 0.5$dB) and stimulated Raman scattering. These
approximations provide quantitative agreement between experiments and
numerical simulations at moderate powers ($<2$W).  Additionnal
numerical simulations show that stimulated Raman scattering has to be
taken into account in order to reproduce very precisely the
experimental PDFs and optical spectra at high values of the mean power
({\it i.e.} $\langle P \rangle=4$W).  However the main physical
results (formation of RWs, emergence of breather-like structures and
heavy-tailed PDFs) are not affected by stimulated Raman scattering.

The random complex field $\psi(t,z=0)$ used as initial condition in
numerical simulations is made from a discrete sum of Fourier components
: 
\begin{equation}\label{ini_field}
 \psi(z=0,t)=\sum\limits_{m} \widehat{X_{m}} e^{i m \Delta \omega t}.
\end{equation}
with $\widehat{X_{m}}=\frac{1}{\Delta T} \int_0^{\Delta T} \psi(z=0,t) e^{-i m \Delta \omega t} dt$ and
$\Delta \omega=2 \pi/\Delta T$. The Fourier modes
$\widehat{X_{m}}=|\widehat{X_{m}}|e^{i \phi_{m}}$ are complex
variables. We have used the so-called random phase (RP) model in which
only the phases $\phi_{m}$ of the Fourier modes are considered as
being random \cite{Nazarenko}. In this model, the phase of each
Fourier mode is randomly and uniformly distributed between $-\pi$ and
$\pi$. Moreover, the phases of separate Fourier modes are not
correlated so that $<e^{i\phi_{n}}e^{i\phi_{m}}>= \delta_{nm}$ where
$\delta_{nm}$ is the Kronecker symbol ($\delta_{nm}=0$ if $n\ne m$ and
$\delta_{nm}=1$ if $n=m$). With the assumptions of the RP model above
described, the statistics of the initial field is stationary, which
means that all statistical moments of the complex field $\psi(z=0,t)$ do
not depend on $x$ \cite{Picozzi:14}. In the RP model, the power
spectrum $n_0(\omega)$ of the random field $\psi(z=0,t)$ reads as : 
\begin{equation}\label{power_spectrum}
<\widehat{X_{n}}\widehat{X_{m}}>=n_{0n} \, \delta_{nm}=n_0(\omega_n). 
\end{equation}
with $\omega_n=n\, \Delta \omega$. In our simulations, we have taken a random complex
field $\psi(z=0,t)$ having a Gaussian optical power spectrum that reads 
\begin{equation}\label{gaussian_ci}
n_0(\omega)=n_0 \, \exp \left[- \left( \frac{\omega^2}{\Delta \omega^2} \right)
 \right]
\end{equation}
where $\Delta \omega=2 \pi \Delta \nu$ is the half width at $1/e$ of the power spectrum.
Statistical properties of the random wave have been computed from Monte Carlo simulation 
made with an ensemble of $10^5$ realizations of the random initial condition.

\section*{Acknowledgments}

This work was supported by the Labex CEMPI (ANR-11-LABX-0007-01) and
by the French National Research Agency (ANR-12-BS04-0011 OPTIROC) and
the BQR \'Emergence-Innovation of Lille~1 University. The authors are
grateful to Francois Anquez and the Biophysics of Cellular Stress
Response group of the PhLAM for the fruitful discussions, their
crucial help, and for providing the sCMOS Camera. The authors are also
grateful to Arnaud Mussot, R\'emi Habert and the photonics group of
the PhLAM for fruitful discussions, for the equipments (the ps laser),
and for the measurement of the GVD of the fiber. The authors thank
Nunzia Savoia for the everyday work on the femto laser and Marc Le
Parquier for his crucial contribution in the development of
the time-lens.

\end{document}